\documentstyle[12pt]{article}
\def\mysection#1{\refstepcounter{section}\subsection{#1}}

\def\thefigure{\thefigure.\arabic{equation}}

\def\ns{\normalsize}
\def\ssize{\normalsize}
\def\l{\left}
\def\r{\right}
\def\un{{1\kern-.25em{\rm l}}}
\def\Bbb#1{{#1\kern-.647em{\rm #1}}}
\def\Z{{\Bbb Z}}
\def\C{{\Bbb C}}
\def\oti{\otimes}
\def\tri{\triangle}
\def\veps{\varepsilon}
\def\del{\delta}
\def\al{\alpha}
\def\be{\beta}

\def\chR{\check{R}}
\def\und#1{{\underline{#1}}}

\newcommand{\nn}{\nonumber}

\begin{document}


\rightline{DAMTP/94-64}
\rightline{NIKHEF-H 94-25}
\vskip 1.5  true cm
\begin{center}
\Large{\bf
Anyonic FRT construction}
\footnote{Talk given at the third colloquium on Quantum Groups and
Physics, Prague, June 1994
}
\\
\vspace{0.45in}
\ns\sc
Shahn Majid
\footnote{Royal Society University Research Fellow and Fellow of
Pembroke College, Cambridge}\\
\ssize\em
Department of Applied Mathematics and
Theoretical Physics\\
University of Cambridge, Silver Street\\
Cambridge CB3 9EW, UK\\
\vspace{0.2in}
\ns\sc
M.J. Rodr\'\i guez-Plaza
\footnote{CSIC (Spain) Research Fellow}\\
\ssize\em
NIKHEF-H\\
Postbus 41882\\
1009 DB Amsterdam, The Netherlands\\
\vspace{0.2in}
\end{center}

\vglue 4 true cm

{\leftskip=1.5 true cm \rightskip=1.5 true cm
\openup 1\jot
{\centerline{\bf Abstract}}
\vspace{0.2in}
\noindent
The Faddeev-Reshetikhin-Takhtajan method to construct matrix bialgebras
from non-singular solutions of the quantum Yang-Baxter equation
is extended to the anyonic or $\Z_n$-graded case. The resulting anyonic
quantum matrices are braided groups in which the braiding is given
by a phase factor.
\\
\par}


\newpage
\openup 2\jot
\mysection{Introduction}

Braided groups \cite{Ma:exa} are a generalization of both quantum
groups and supergroups.
They have a coproduct $\und\Delta$ which is a homomorphism
not to the usual commutative tensor product but to the braided
tensor product.
The two factors in the braided tensor product no longer
commute but instead have braid statistics described by a braiding $\Psi$.
A superquantum group has $\Psi$ given by super-transposition (with a factor
$\pm 1$). The next simplest and first truly braided case is
the notion of an anyonic quantum group \cite{Ma:any} \cite{Ma:csta}
\cite{MaPla:ran} where $\Psi$ has
a factor $e^{2\pi i/n}$ in place of $\pm 1$.

This note is a short announcement of \cite{MRP2} where
the construction of quantum matrices of this anyonic type
is achieved. We recall that the Faddeev-Reshetikhin-Takhtajan
(FRT) construction \cite{FRTlie} associates a
quantum matrix bialgebra $A(R)$ to each
non-singular solution $R$ of the quantum Yang-Baxter equation.
Its matrix of generators $\{t^i{}_j\}$ is like the co-ordinates on
the set of matrices $M_n$, i.e. these bialgebras are of `function
algebra type'.
There is not exactly in \cite{FRTlie} a corresponding construction
for the dual bialgebras of `enveloping algebra type' but there is
a dually-paired bialgebra $\widetilde{U(R)}$ with matrix generators
$l^\pm$ which can generally be quotiented down to something like
an enveloping algebra quantum group.
A nice thing about this approach is that the bialgebras and also the
duality pairing have as input data only the $R$-matrix.
Many R-matrices are known so we have a rich and powerful construction.
The generalization of this FRT or R-matrix setup to
the case of super-quantum matrices is also known and is equally
rich \cite{Liao} \cite{MRP}. The input data in this case is
a solution of the {\em super
Yang-Baxter equation}, which is still older \cite{KulSkl}. Our paper
\cite{MRP} gave an abstract picture of this super FRT procedure in terms of a
theory of transmutation which can be used to construct super and braided
quantum groups from ordinary ones. Its extension now in the anyonic matrix case
is one of our motivations.

By an anyonic or $\Z_n$-graded bialgebra we understand more precisely that
the algebra structure is $\Z_n$-graded and that two copies of the
algebra are multiplied according to the rule
\begin{equation}
\l(a\oti b\r)\l(c\oti d\r)=e^{{{2\pi i}\over n}p(b)p(c)}\l(ac\oti bd\r)
\label{braiding}
\end{equation}
where $a,b,c,d$ are homogeneous elements in the algebra whose grading
is given by the function $p(\cdot)$ with values in the set
$\Z_n=\{0,1,\ldots,n-1\,\, ({\rm mod}\,n)\}$. The coalgebra is defined with
coproduct
$\und\Delta$ which is an algebra homomorphism from one copy to two copies
multiplied in this way.
We will obtain anyonic bialgebras $\und A(\und R)$ with a matrix $\{a^i{}_j\}$
of generators starting
form the {\em $\Z_n$-graded Yang-Baxter equation}, which we also introduce. The
$n=1$ case recovers
the usual or ungraded quantum matrices, while the $n=2$ case recovers the
super-quantum matrices mentioned above.
Also in \cite{MRP2} is the theory of anyonic quantum groups
$\und U(\und R)$ of enveloping algebra type,
as well as anyonic quantum planes $\und V(\und R)$ on which they act.
We shall outline some of this further theory too in the
last section of this note.

It should be stressed that the jump from $n=2$,
the {\em super} case, to $n>2$ gradation
represents a drastic discontinuity due to the fact that many
properties that hold for the former do not always hold in an analogous manner
for the latter. This is because for $n>2$  the permutation group generated by
transpositions $\Psi$ must truly be replaced by the Artin braid group. One of
the resulting complications is manifested in the following example:
a $\Z_n$-graded algebra that satisfies the relation
$a \, b=e^{{{2\pi i}\over n}p(a)p(b)}\, b \, a$
for all elements $a$ and $b$ in the
algebra is called super-commutative when $n=2$ but when
$n>2$ the relation is inconsistent; it is obvious then that whatever
the definition of anyonic-commutativity should be in the $n>2$
case it cannot be given by such formula.
Another example is provided by the Jacobi
identity for Lie superalgebras. Since there is such an identity it is not
unreasonable to wonder about its $\Z_n$-graded analogue.
We think that these unknowns can be best answered by first building up a
good
supply of anyonic quantum groups. Anyonic quantum matrices of function algebra
type should
give us some insight into the correct notion of anyonic-commutative algebras,
while
their duals of enveloping algebra type should give us some insight into the
correct notion
of anyonic Lie algebras, in both cases as the `classical limit' of suitable
$\Z_n$-graded
examples. The notion
of `quantization' or q-deformation is independent of the notion of anyonic or
other braiding.
This can be regarded as one of the general motivations behind \cite{MRP2}.

Other applications of the graded FRT bialgebras would be to
physical systems where the relation (\ref{braiding}) is involved.
Of course this is not the case of a 4-dimensional world where all the
operators associated to real particles satisfy either commuting or
anticommuting relations (they are operators in an ungraded or in a super
algebra) depending whether they describe bosons or fermions but it
might be a 2-dimensional one where some anyonic behaviour is described.
This is also the reason why we refer to $\Z_n$-gradation when
$n>2$ as {\em anyonic}.  Moreover, in any dimension the process of
$q$-deformation (for
example for the purpose of $q$-regularization or as a quantum correction to
geometry) also spoils the
spin-statistics theorem and indeed does typically induce braid
statistics. For example, the dilaton generator in some approaches
to the q-deformed Poincar\'e group reflects a residual $\Z$-grading
(with corresponding braid statistics)
induced by the $q$-deformation.

To conclude this introduction, we mention that the ultimate
justification for studying the anyonic bialgebras derived from the
modified FRT construction that we expose is that this is a
construction simple to manipulate and without inconsistencies.
There is no doubt then that these anyonic structures can teach
something to us.

\mysection{Anyonic Yang-Baxter equation}

As we have mentioned in the introduction the consideration of
gradation brings some modifications with respect to the ungraded
situation. The first modification is that the ordinary
Yang-Baxter equation is substituted by its graded analogue. In this
case of $\Z_n$-gradation that we are studying its graded analogue
is referred as the {\em anyonic Yang-Baxter equation} and
it is introduced in the present section.

Let us consider a vector space $V$ of finite dimension $d$ with
the direct sum decomposition $V=\oplus_{\al=0}^{n-1} V_\al, \, n\ge 2$
such that there is defined a function with values in the group $\Z_n$ given
by $p(x)\equiv\al \,\, ({\rm mod}\,n)$ when $x$ lies in $V_\al$.
This direct sum defines a $\Z_n$-gradation on $V$ and any vector
$v\in V$ is uniquely determined by its {\em homogeneous component of
degree $\al$} $v_\al\in V_\al$ by way of the expansion
$v=\oplus_{\al=0}^{n-1} v_\al.$
The gradation extends to the vector space
$V\oti V=\oplus_{\al, \be=0}^{n-1} V_\al\oti V_\be$ which also becomes
$\Z_n$-graded with the degree of any
homogeneous element $v_\al\oti v_\be$ in $V_\al\oti V_\be$ defined by
$p\l(v_\al\oti v_\be\r)\equiv p\l(v_\al\r)+p\l(v_\be\r)=
\al+\be \,\, ({\rm mod}\,n)$.

Let $\{e_i\oti ej\}, i,\, j \in I$ be a linear basis for $V\oti V$
with the indexing set $I\equiv\{1,\ldots, d\}$ and assume that all
vectors $e_i$ are homogeneous of degree $p(e_i)=p(i)$ where
we use the convention that Latin indices run over the set $I$
to avoid confusion with Greek indices used to denote degree.

Under these assumptions we say that an invertible matrix $\und R$
in ${\rm End}\l(V\otimes V\r)$  is an {\em anyonic R-matrix} if
\begin{eqnarray}
\und R^a{}_c{}^b{}_d=0\quad {\rm when}
\quad p(a)+p(b)-p(c)-p(d)\ne 0 \,\, ({\rm mod}\, n),
\label{null}
\end{eqnarray}
which we call the `null degree condition' and
\begin{eqnarray}
e^{{{2\pi i}\over n}p(e)\l[p(f)-p(c)\r]}
\,\und R^b{}_e{}^a{}_f\,\und R^i{}_k{}^f{}_c\,\und R^k{}_j{}^e{}_d=
e^{{{2\pi i}\over n}p(r)\l[p(s)-p(a)\r]}
\,\und R^i{}_p{}^b{}_r\,\und R^p{}_j{}^a{}_s\,\und R^r{}_d{}^s{}_c,
\label{AYBE}
\end{eqnarray}
which we call the {\em anyonic Yang-Baxter equation}. A
sum over repeated indices is understood.
Obviously the ungraded Yang-Baxter equation is obtained from this one
with the particularization $n=1$. Likewise, the
super Yang-Baxter equation of \cite{KulSkl} is recovered with
the choice $n=2$.

Note that for $n=1$ the condition (\ref{null}) is empty since
any integer is zero mod 1. This is why we do not see it in the
usual FRT procedure. For $n=2$ the condition is equivalent to
saying that $\und R$ is an even matrix since the degree
of each matrix element is given by
$p\l(\und R^a{}_c{}^b{}_d\r)=p(a)+p(b)-p(c)-p(d)$ for all $n$.

There is no big loss of generality in adopting this null degree
assumption and it has, on the contrary, many advantages for the
simplifications that it introduces in the calculations.
Neither this condition (\ref{null}) nor the relation
(\ref{AYBE}) change under a basis transformation in $V$ of the type
$e'_j=S_j{}^i e_i$ with $p(e'_i)= p(e_i)$ for all $i$ and
thus both conditions are invariant under the transformation
$\und R\rightarrow \l(S\oti S\r)^{-1}\,\und R\l(S\oti S\r)$
where $S$ is any non-singular $d^2\times d^2$  matrix of null degree.

The following properties characterize anyonic R-matrices:
\begin{enumerate}
\item A solution to the {\em braid relation}
\begin{eqnarray}
\chR^a{}_e{}^b{}_f\,\chR^f{}_k{}^i{}_c\,\chR^e{}_j{}^k{}_d=
\chR^b{}_p{}^i{}_r\,\chR^a{}_j{}^p{}_s\,\chR^s{}_d{}^r{}_c
\label{braid}
\end{eqnarray}
can be obtained from every solution of (\ref{null})--(\ref{AYBE})
through the transformation
$\chR=\und P\,\und R$ where $\und P$ denotes the
{\em anyonic permutation operator} in the $\Z_n$-graded vector space $V\oti V$
defined by its action
$\und P\l(a\oti b\r)=e^{{{2\pi i}\over n}p(a)p(b)}\, b\oti a$
on homogeneous vectors $a$ and $b$ in $V$ or by its components
on any linear basis of $V$ (we assume basis transformation in $V$ as
those mentioned above)  $\und P^a{}_c{}^b{}_d=e^{{{2\pi i}\over
n}p(a)p(b)}\,\del^a_d\del^b_c$. A simple calculation indicates that this
operator represents the $n$th-root of the permutation operator $P$
with action $P\l(a\oti b\r)=b\oti a$ when $n$ is odd and the
$n$th-root of the identity operator in $V\oti V$ when $n$ is even.

Note that the equality (\ref{braid}) does not contain any degree label
$p(\cdot)$ which shows that the gradation does not alter the braid
relation.

\item There exists a relation between the solutions of
(\ref{null})--(\ref{AYBE}) and the solutions of the ungraded
Yang-Baxter equation given by the formula
\begin{equation}
R^a{}_c{}^b{}_d=e^{{{2\pi i}\over n}p(a)p(b)} \und R^a{}_c{}^b{}_d,
\label{relation}
\end{equation}
where the non underlined variables refer to the ungraded situation.

\item The matrices $\und R^\pm$ defined by
$\und R^+\equiv\und P\,\und R\,\und P$ and $\und R^-\equiv{\und R}^{-1}$
satisfy both the same equation
\begin{eqnarray}
e^{{{2\pi i}\over n}p(e)\l[p(c)-p(f)\r]}
\,{{\und R}^\pm}^b{}_e{}^a{}_f\,{{\und R}^\pm}^i{}_k{}^f{}_c
\,{{\und R}^\pm}^k{}_j{}^e{}_d=
e^{{{2\pi i}\over n}p(r)\l[p(a)-p(s)\r]}
\,{{\und R}^\pm}^i{}_p{}^b{}_r
\,{{\und R}^\pm}^p{}_j{}^a{}_s\,{{\und R}^\pm}^r{}_d{}^s{}_c,
\label{plus}
\end{eqnarray}
but this does {\em not} coincide with equation (\ref{AYBE}) except in the
ungraded or super case.
\end{enumerate}

The proof of all these properties uses the null degree condition
satisfied by the matrix elements of $\und R$. Properties 1 and 2 are
very helpful to find solutions to the anyonic Yang-Baxter
equation since, for instance, many known solutions of the
ordinary Yang-Baxter relation are transformable into solutions of
its anyonic counterpart.
Of course one should be careful and be sure that in the transformation
the requisite of null parity of $\und R$ holds. An application of
property 3 will be mentioned in the last section.

\mysection{Anyonic FRT construction}

We know from \cite{FRTlie} that to any non-singular $d^2\times d^2$
matrix solution $R$ of the ordinary Yang-Baxter equation (we dont use
underlines when we refer to the ungraded case) it is possible to
associate a matrix bialgebra $A\l(R\r)$ generated by $\un$ and
$\l\{t^i{}_{j}\r\}\, i, j=1,\ldots,d$ with algebra relations
\begin{equation}
R^a{}_f{}^b{}_e\, t^f{}_{c}\,t^e{}_{d}=
t^b{}_{r}\,t^a{}_{s}\,R^s{}_c{}^r{}_d,
\label{FRTal}
\end{equation}
and coalgebra given by
\begin{equation}
\tri t^i{}_{j}=\sum_{k=1}^d t^i{}_{k}\oti t^k{}_{j},\quad
\veps\l(t^i{}_j\r)=\delta^i{}_j.
\label{FRTcoal}
\end{equation}

As we have explained in the introduction the association is still possible in
the $\Z_n$-graded case but to a solution of (\ref{null})--(\ref{AYBE})
instead. Thus we define the associated anyonic
bialgebra,  $\und A\l(\und R\r)$, as generated by $\un$ and
$\l\{a^i{}_{j}\r\}\, i, j=1,\ldots,d$ with algebra and coalgebra
relations
\begin{eqnarray}
&e^{{{2\pi i}\over n}\l[p(a)p(b)+p(c)p(e)\r]}
\,\und R^a{}_f{}^b{}_e\, a^f{}_{c}\,a^e{}_{d}
=e^{{{2\pi i}\over n}\l[p(c)p(d)+p(r)p(a)\r]}
\,a^b{}_{r}\,a^a{}_{s}\,\und R^s{}_c{}^r{}_d,
&\label{anyFRTal}\\
&\und\tri a^i{}_{j}=\sum_{k=1}^d a^i{}_{k}\oti a^k{}_{j},\quad
\und\veps\l(a^i{}_j\r)=\delta^i{}_j.
&\label{anyFRTcoal}
\end{eqnarray}
The algebra structure of this new bialgebra is $\Z_n$-graded with the
generators of the bialgebra defined as homogeneous elements of
degree $p\l(a^i{}_{j}\r)\equiv p(i)-p(j)\,\,({\rm mod}\,n)$
and as a graded algebra the product of any
two homogeneous elements $x,y$ in the algebra is homogeneous and
moreover satisfies $p(xy)=p(x)+p(y)$.

Notice that it is not possible to get rid of the phases in
equation (\ref{anyFRTal}) with the transformation
(\ref{relation}). This is an indication of the intrinsic anyonic
character of the algebra that we are introducing and shows that the graded
algebra is {\em not} the ungraded one (\ref{FRTal}) with the
substitution of the ordinary Yang-Baxter equation solution by its graded
analogue. Rather, one is obtained from the other by a process of transmutation,
which changes the algebra product.

Regarding the coalgebra structure (\ref{anyFRTcoal})
$\und \tri a^i{}_{j}$ satisfies the
relation (\ref{anyFRTal}) provided that two copies of the algebra
multiply according to the law
\[
\l(a\oti b\r)\l(c\oti d\r)=e^{{{2\pi i}\over n}p(b)p(c)}\l(ac\oti bd\r)
\]
for all homogeneous elements $a,b,c,d$ in $\und A\l(\und R\r)$. This
law corresponds to the particular case $\Psi\l(b\oti
c\r)=e^{{{2\pi i}\over n}p(b)p(c)}\, c\oti b$
of a more general concept of
{\em braiding} $\Psi$ \cite{Ma:exa}. Although it is not our purpose to review
braided groups here, we see that the construction presented in this
note modifies the FRT approach to also include the anyonic braiding.

We now give a non-trivial example of our construction, for
$n=3$. Our starting point is the solution of the ordinary Yang-Baxter equation
given by
\begin{eqnarray}
R=\l(\begin{array}{ccccccccc}
1&0&0&0&0&0&0&0&0\\
0&1&0&0&0&0&0&0&0\\
0&0&1&0&1-q&0&0&0&0\\
0&0&0&1&0&0&0&0&0\\
0&0&0&0&q&0&0&0&0\\
0&0&0&0&0&q^2&0&q-1&0\\
0&0&0&0&0&0&1&0&0\\
0&0&0&0&0&0&0&q^2&0\\
0&0&0&0&0&0&0&0&q
\label{exR}
\end{array}\r)
\end{eqnarray}
taken from section 4 of \cite{Ma:sol} and written in the basis
$\{e_1\oti e_1, e_1\oti e_2,\ldots,e_3\oti e_3\}$ of $V\oti V$.
Here $q=e^{2\pi i/3}$. Let us assume that $V$ is a $\Z_3$-graded
vector space with degree of each basis vector given by
$p(e_i)=i-1$ for all $i=1,2,3$ (note that the obvious inequality
$d\ge n$ holds then). Under these assumptions this solution is associated by
(\ref{relation}) with the solution of the anyonic Yang-Baxter
equation with zero entries except in the places
\[
\und R^i{}_i{}^j{}_j=1\quad i,j=1,2,3\qquad \und R^1{}_2{}^3{}_2=1-q
\qquad \und R^2{}_3{}^3{}_2=q^2-q
\]
where the left indices here numbered the block.
Obviously this is a null degree solution with respect to the previous
degree assignation $p(e_i)$. In the
anyonic bialgebra (\ref{anyFRTal})--(\ref{anyFRTcoal}) the generators
$a^i{}_{i}$ are of degree zero, $a^1{}_{3},a^2{}_{1},a^3{}_{2}$ are
of degree $1$ and $a^1{}_{2},a^2{}_{3},a^3{}_{1}$ are of degree $2$
and satisfy the following relations: $a^1{}_{1}$ is central and
\begin{eqnarray}
\begin{array}{lll}
a^3{}_{3} a^3{}_{2}=a^3{}_{2} a^3{}_{3} & a^3{}_{3} a^3{}_{1}=0& a^3{}_{3}
a^2{}_{3}=qa^2{}_{3}
a^3{}_{3}
\nn\\
a^3{}_{3} a^2{}_{2}=a^2{}_{2} a^3{}_{3}+(q-1) a^2{}_{3} a^3{}_{2}
& a^3{}_{3} a^2{}_{1}=q^2 a^2{}_{1} a^3{}_{3} & a^3{}_{3} a^1{}_{3}=q^2
a^1{}_{3} a^3{}_{3}\nn\\
a^3{}_{3} a^1{}_{2}=q a^1{}_{2} a^3{}_{3}+q (q-1) a^1{}_{3} a^3{}_{2}
& a^3{}_{2} a^3{}_{1}=0& a^3{}_{2} a^2{}_{3}=a^2{}_{3} a^3{}_{2}\nn\\
a^3{}_{2} a^2{}_{2}=a^2{}_{2} a^3{}_{2}+q(q-1) a^2{}_{1} a^3{}_{3}
& a^3{}_{2} a^2{}_{1}=a^2{}_{1} a^3{}_{2}& a^3{}_{2} a^1{}_{3}=a^1{}_{3}
a^3{}_{2}\nn\\
a^3{}_{2} a^1{}_{2}=a^1{}_{2} a^3{}_{2}+ (q^2-1) a^1{}_{3} a^3{}_{1}& a^3{}_{1}
a^3{}_{3}=0
& a^3{}_{1} a^3{}_{2}=0\nn\\
(a^3{}_{1})^2=0& a^3{}_{1} a^2{}_{3}=q^2 a^2{}_{3} a^3{}_{1}& a^3{}_{1}
a^2{}_{2}=a^2{}_{2} a^3{}_{1}\nn\\
a^3{}_{1} a^2{}_{1}=0& a^3{}_{1} a^1{}_{3}=q a^1{}_{3} a^3{}_{1}&
a^3{}_{1} a^1{}_{2}=q^2 a^1{}_{2} a^3{}_{1}\nn\\
(a^2{}_{3})^2=q a^1{}_{3} a^3{}_{3}& a^2{}_{3} a^2{}_{2}=q a^2{}_{2} a^2{}_{3}&
a^2{}_{3} a^2{}_{1}=0\nn\\
a^2{}_{3} a^1{}_{3}=0& a^2{}_{3} a^1{}_{2}=q^2 a^1{}_{2} a^2{}_{3}
&(a^2{}_{2})^2=a^1{}_{1} a^3{}_{3}+ a^1{}_{2} a^3{}_{2}+q^2 a^1{}_{3}
a^3{}_{1}\nn\\
a^2{}_{2} a^2{}_{1}=0& a^2{}_{2} a^1{}_{3}=q a^1{}_{3} a^2{}_{2}
& a^2{}_{2} a^1{}_{2}=q^2 a^1{}_{2} a^2{}_{2}+(1-q) a^1{}_{3} a^2{}_{1}\nn\\
a^2{}_{1} a^3{}_{1}=0& a^2{}_{1} a^2{}_{3}=0& a^2{}_{1} a^2{}_{2}=0\nn\\
(a^2{}_{1})^2=0& a^2{}_{1} a^1{}_{3}=q^2 a^1{}_{3} a^2{}_{1}
& a^2{}_{1} a^1{}_{2}=q a^1{}_{2} a^2{}_{1}\nn\\
a^1{}_{3} a^2{}_{3}=0 &(a^1{}_{3})^2=0&a^1{}_{3} a^1{}_{2}=0 \nn\\
a^1{}_{2} a^1{}_{3}=0&(a^1{}_{2})^2=q a^1{}_{1} a^1{}_{3}&\nn
\end{array}
\end{eqnarray}
together with the ``strange'' relations
\begin{eqnarray}
\begin{array}{ll}
a^1{}_{3} a^2{}_{2}=-a^1{}_{2} a^2{}_{3} &a^2{}_{2} a^3{}_{1}=-q a^2{}_{1}
a^3{}_{2}\nn\\
a^2{}_{3} a^3{}_{1}=-q^2 a^2{}_{1} a^3{}_{3}
&a^1{}_{3} a^2{}_{1}=-a^1{}_{1} a^2{}_{3}-q^2 a^1{}_{2} a^2{}_{2}\nn\\
a^2{}_{2} a^2{}_{3}=-q a^1{}_{2} a^3{}_{3}-q^2 a^1{}_{3} a^3{}_{2}.&\nn
\end{array}
\end{eqnarray}
{} From these formulae and the fact that $q$ is a primitive
root of unity it is simple to derive that
$a^1{}_{2},a^1{}_{3},a^2{}_{1},a^2{}_{3},a^3{}_{1}$
have cube equal to zero and that the cube of
$a^2{}_{2}, a^3{}_{2}, a^3{}_{3}$ are all central.
It is interesting to realize that this algebra can be considerably
reduced by quotienting it by the relations
$a^1{}_{1}=1,\,\,a^2{}_{1}=a^3{}_{1}=0$. These are all compatible with the
coalgebra structure. After this reduction the anyonic bialgebra
above can be expressed in terms of only three independent
generators, namely $z_1\equiv a^2{}_{2},\,z_2\equiv a^3{}_{2},
\,z_3\equiv a^1{}_{2}$ of degree $0,1,2$ respectively
(the remaining generators being expressed in terms of them by
$a^1{}_{3}= q^2 z_3^2$, $a^2{}_{3}= -q^2 z_3 z_1$,
$a^3{}_{3}=z_1^2-z_3 z_2$ using the relations above).
The quotient bialgebra is generated by these $z_1,z_2,z_3$
with algebra and coalgebra relations
\begin{equation}
z_2 \,\,{\rm central}\qquad z_1z_3=q^2z_3z_1\qquad z_3^2=0
\label{alzzz}
\end{equation}
\begin{eqnarray}
&&\und\tri z_1=z_1\oti z_1-z_1z_3\oti z_2\nn\\
&&\und\tri z_2=z_2\oti z_1+z_1^2\oti z_2-z_3z_2\oti z_2
\label{coalzzz}\\
&&\und\tri z_3=\un\oti z_3+z_3\oti z_1+q^2 z_3^2\oti z_2\nn\\
&& \und\veps\l(z_1\r)=1, \quad \und\veps\l(z_2\r)=\und\veps\l(z_3\r)=0.\nn
\end{eqnarray}
Notice that the coalgebra stated implies that the central element
$z_1^3$ is group-like. In this particular example it is possible to
extend the anyonic bialgebra to an anyonic Hopf algebra just by assuming
that the operator $z_1^{-1}$ exists. With this condition
the anyonic-antipode $\und S$ of each generator is well defined and given by
\[
\und S\l(z_1\r)=z_1^{-1}-z_1^{-3}z_2z_3\qquad
\und S\l(z_2\r)=-z_1^{-3}z_2\qquad
\und S\l(z_3\r)=-z_3z_1^{-1}-q z_1^{-3}z_2z_3^2.
\]
Any element of this graded Hopf algebra is expressed as a product on
the generators and its antipode is calculated with the property
$\und S\l(ab\r)=q^{p(a)p(b)}\und S\l(b\r)\und S\l(a\r)$ in terms of
antipode of the generators. We recall again here that $q=e^{2\pi i/3}$.

This example demonstrates a solution to one of the puzzles underlying any naive
approach to the anyonic theory: we might well expect in many anyonic or
$\Z_n$-graded algebras to have some generators with the $n$th power equal to
zero. This could be viewed as a partial substitute for anyonic
commutativity (just as super-commutativity implies that the
square of an odd generator is zero). But the puzzle is how such cubic or higher
order
relations could come from an FRT-type construction where the relations are
quadratic. We see now how this works in our example
for $n=3$ where, for example,  $a^1{}_2$ has square equal to a product of two
other generators one of which is zero when multiplied by another $a^1{}_2$.
Starting with quadratic relations even for arbitrary $n$
is not an obstacle since the prescription to formulate anyonic bialgebras
that we have presented here, in spite of its quadratic aspect, gives the right
anyonic character to the bialgebra that it constructs.

\mysection{Final remark}

The material presented in this paper details one of the two graded
bialgebras afforded by the modified FRT construction.
The second one of universal enveloping algebra type will be fully
studied in \cite{MRP2}. However, any of the two can be obtained from
the other once the pairing between them is known so
we anticipate the pairing here. Thus the anyonic bialgebra
$\und A\l(\und R\r)$ of the previous section is dually paired with
another anyonic bialgebra $\und U\l(\und R\r)$ with generators
$1,m^\pm{}^i_j,\,\,i,j=1,\ldots,d$ and the pairing is given by
\[
<m^\pm{}^i_j, a^k{}_{l}>=e^{{{2\pi i}\over n}p(i)\l[p(k)-p(l)\r]}
\,{{\und R}^\pm}^i{}_j{}^k{}_l,
\]
where ${\und R}^\pm$ indicates a solution of equation (\ref{plus}).

%
\def\section{\subsection}

\end{document}